\documentclass[a4paper]{article}
\usepackage{INTERSPEECH2022}

\usepackage{bm}
\usepackage{graphicx}
\usepackage{cite}
\usepackage{amsmath,amssymb,amsfonts,mathrsfs}
\usepackage{multirow}
\usepackage{graphicx}
\usepackage{amsxtra}
\usepackage{threeparttable}
\usepackage{cite}
\usepackage{mathrsfs}
\usepackage{algorithm,algorithmic}

\usepackage{multirow, booktabs}
\usepackage{graphicx}
\usepackage{textcomp}
\usepackage{xcolor}
\usepackage{url}

\DeclareMathOperator*{\argmax}{argmax}
\DeclareMathOperator{\sinc}{sinc}

\title{Joint Optimization of Sampling Rate Offsets Based on \\
Entire Signal Relationship Among Distributed Microphones}
\name{Yoshiki Masuyama$^1$, Kouei Yamaoka$^1$, Nobutaka Ono$^1$}
\address{$^1$Tokyo Metropolitan University, Tokyo, Japan}
\email{masuyama-yoshiki@ed.tmu.ac.jp}

\begin{document}
\maketitle
\begin{abstract}
In this paper, we propose to simultaneously estimate all the sampling rate offsets (SROs) of multiple devices.
In a distributed microphone array, the SRO is inevitable, which deteriorates the performance of array signal processing.
Most of the existing SRO estimation methods focused on synchronizing two microphones.
When synchronizing more than two microphones, we select one reference microphone and estimate the SRO of each non-reference microphone independently.
Hence, the relationship among signals observed by non-reference microphones is not considered.
To address this problem, the proposed method jointly optimizes all SROs based on a probabilistic model of a multichannel signal.
The SROs and model parameters are alternately updated to increase the log-likelihood based on an auxiliary function.
The effectiveness of the proposed method is validated on mixtures of various numbers of speakers.
\end{abstract}


\section{Introduction}

Microphone array signal processing, including blind source separation (BSS)~\cite{bss1,bss2}, is a fundamental technique with various applications such as automatic speech recognition~\cite{bssasr2} and sound event detection~\cite{bsssed}.
Although More microphones are desirable to improve the performance of BSS~\cite{overiva1,overiva2}, it is costly to prepare a large microphone array.
This limits the number of microphones and the array size in many applications.
To address this problem, distributed microphone array (DMA) processing has gained considerable attention~\cite{dma1,dma2}.
A DMA exploits a set of microphones on multiple devices, including tablets and smartphones, and does not require any specialized devices.
By using a DMA, we can acquire a large number of observations and conduct array signal processing such as sound source localization~\cite{dmassl2}, speech enhancement~\cite{dmase1,dmase2}, and BSS~\cite{dmabss2}.

In a DMA, microphones are connected to device-dependent analog-to-digital converters, and their sampling rates are slightly different even when the nominal ones are the same.
These sampling rate offsets (SROs) deteriorate the performance of array signal processing including BSS~\cite{ml1}.
We should thus estimate and compensate for the SROs in advance~\cite{ml1,ml2,cd1,cd2,cd3,cm2,dxcp1,dxcp2,auxdxcp,rbi,srossl,sro-music,overlapsave}.
Since array signal processing is often conducted in the time-frequency (T-F) domain, an SRO model in the T-F domain, called the linear phase drift (LPD) model, has been widely used~\cite{cd1}.
The LPD model considers that the SRO changes the phase of the short-time Fourier transform (STFT) coefficients linearly with time and frequency.
Various SRO estimation approaches have been developed based on this model~\cite{ml1,ml2,cd1,cd2,cd3,cm2}.

The first approach, called the coherence-drift-based approach~\cite{cd1,cd2,cd3}, computes the complex coherence between the signals observed by the reference and non-reference microphones.
The SRO is estimated from the ratio of the complex coherence between successive time frames.
The second approach relies on a probabilistic model of the two-channel signals~\cite{ml1,ml2}.
This approach assumes that the STFT coefficients of the synchronized signals follow a multivariate complex Gaussian distribution.
On the basis of this assumption, the SRO is estimated in a maximum likelihood manner.
The third approach estimates the SRO to maximize the correlation between the STFT coefficients~\cite{cm2}.
It is shown that the correlation takes the largest value when the SRO is accurately compensated for.
The second and third approaches often require a higher computational cost to search for the optimal SRO, but they have achieved promising results.

\begin{figure}[t]
\centering
\includegraphics[width=0.99\columnwidth]{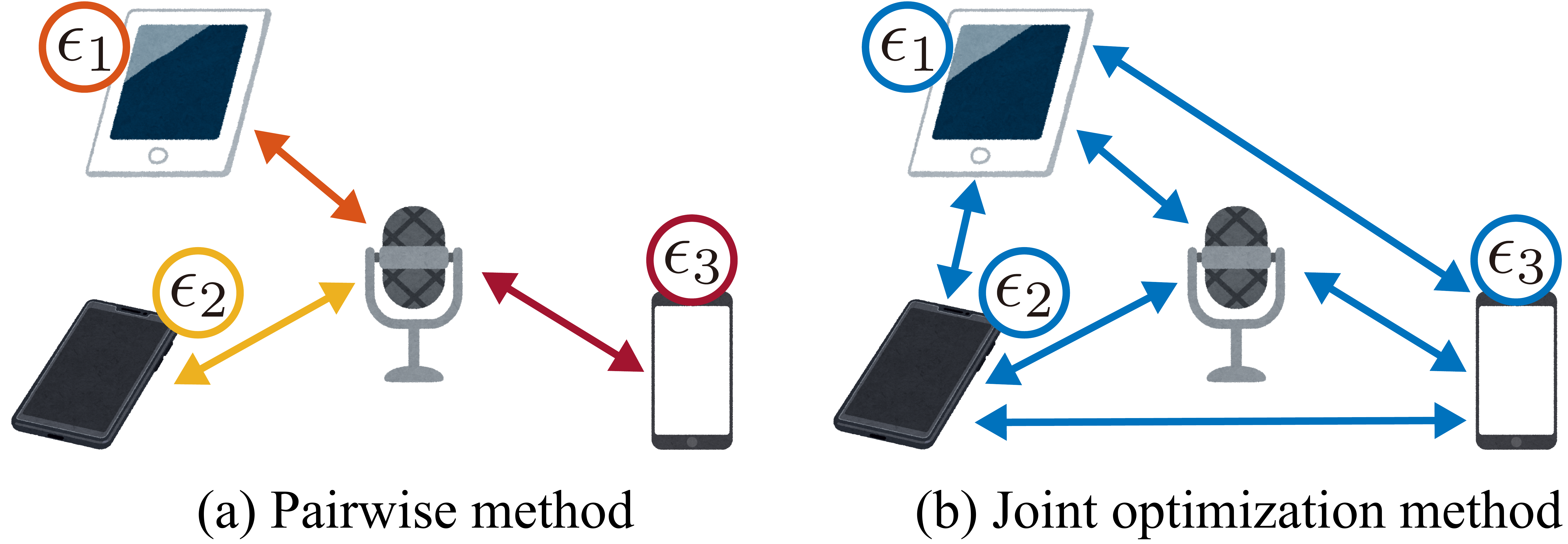}
\caption{Illustration of (a) existing pairwise synchronization and (b) proposed joint-optimization-based synchronization.
As an example, the center microphone is considered as the reference one.
Each color corresponds to an optimization problem.}
\label{fig:overview}
\vspace{-2pt}
\end{figure}

The aforementioned approaches are based on the relationship between the signals observed by one reference microphone and one non-reference microphone~\cite{ml1,ml2,cd1,cd2,cd3,cm2,dxcp1,dxcp2,auxdxcp}.
When synchronizing more than two microphones, we should select the reference microphone and estimate the SRO of each non-reference microphone independently, as depicted in Fig.~\ref{fig:overview}-(a).
We call this type of synchronization the pairwise method.
One of the drawbacks of this method is that its performance depends on the selection of the reference microphone.
This is because the relationship among the non-reference microphones is not taken into account.
Meanwhile, it is still a challenging task to select the optimal reference microphone in terms of synchronization.
It is thus desirable to exploit full spatial information of the signals observed by a DMA regardless of the reference microphone.

To this end, we propose to consider the relationships between all microphone pairs and optimize all the SROs jointly as depicted in Fig.~\ref{fig:overview}-(b).
The proposed method estimates the SROs in a maximum likelihood manner through a probabilistic model of the entire multichannel signal.
Although the golden section search has been used to optimize each SRO independently~\cite{ml1}, it is not directly applicable to the joint optimization of all SROs.
To address this problem, we present an iterative algorithm using an auxiliary function and guarantee the non-decrease property of the log-likelihood in the algorithm.
Our experimental results showed that the proposed joint optimization method outperformed the existing pairwise methods~\cite{ml1,cm2}.

\section{SRO Estimation in T-F Domain}

\subsection{SRO Model in T-F Domain}

Let us assume that an $M$-channel signal is observed by a DMA, where the $0$th microphone is selected as the reference one.
The sampling rate of the $m$th microphone is given by
\begin{equation}
    r_m = (1 + \epsilon_m) r_0,
\end{equation}
where $m = 0, \ldots, M-1$ is the microphone index, $\epsilon_m$ is the SRO of the $m$th microphone, and $\epsilon_0 = 0$.
Let $\widetilde{\chi}_m$ be a continuous signal to be measured by the $m$th microphone.
Then, the $\tau$th entry of its discrete version is given by
\begin{align}
    \chi_{m}[\tau] = \widetilde{\chi}_m \left(\frac{\tau}{(1+\epsilon_m) r_0} + \Delta_m \right),
\end{align}
where $\Delta_m$ is the sampling time offset (STO) of the $m$th microphone.
Although an accurate estimation of the STO is not easy, its small error is acceptable for BSS.
We thus hereafter assume that the STO is already recovered by an existing method~\cite{ml1}.

Since array signal processing is often conducted in the T-F domain owing to its efficiency, the LPD model has been widely used to estimate and compensate for the SRO.
Let the STFT of the discrete signal $\bm{\chi}_m$ with a window $\bm{g}$ of length $L$ be
\begin{equation}
    x_m[t,f] = \sum_{l=0}^{L-1} \chi_m[l + a t] g[l] \mathrm{e}^{- 2 \pi j f l / F},
\end{equation}
where $j$ is the imaginary unit, $a$ is the window shift, and $t=0,\ldots,T-1$ and $f=0,\ldots,F-1$ are the time frame and frequency bin indices, respectively.
The LPD model represents the SRO by a phase modification of STFT coefficients and compensates for it as follows \cite{cd1}:
\begin{equation}
    \widehat{x}_m[t,f] = x_m[t,f] \exp\left(\frac{2 \pi j a t f \epsilon_m}{F} \right).
    \label{eq:resampling}
\end{equation}
That is, $\widehat{x}_m[t,f]$ can be interpreted as the STFT coefficient of the synchronized signal when $\epsilon_m$ is accurately estimated.

\subsection{Maximum Likelihood Estimation of SRO}

Based on the LPD model, the probabilistic-model-based approach estimates the SRO in a maximum likelihood manner~\cite{ml1,ml2}.
In this approach, the compensated STFT coefficients $\widehat{\bm{x}}[t,f] = [\widehat{x}_0[t,f], \ldots, \widehat{x}_{M-1}[t,f]]^\mathsf{T}$ are assumed to follow a multivariate complex Gaussian distribution:
\begin{equation}
    \widehat{\bm{x}}[t,f] \sim \mathcal{N}_\mathbb{C} (\bm{0}, \bm{V}[f]), \label{eq:gauss}
\end{equation}
where $\bm{V}[f]$ is the spatial covariance matrix (SCM) of the synchronized signals.
The probabilistic model in \eqref{eq:gauss} implies that the sound sources do not move and their powers are stationary.

The existing methods have addressed the case $M=2$.
In such a case, the log-likelihood for \eqref{eq:gauss} can be reformulated to the following univariate objective function with respect to $\epsilon_1$~\cite{ml2}:
\begin{align}
    &\mathcal{I}(\epsilon_1) = - \sum_f \log \left( \sum_t |x_0[t,f]|^2 \sum_t |\widehat{x}_1[t,f]|^2 \right. \nonumber \\
    &\hspace{105pt} \left. - \Bigl| \sum_t x_0[t,f] \widehat{x}_1[t,f] \Bigr|^2 \right),
    \label{eq:objective-2ch}
\end{align}
where $\widehat{x}_1[t,f]$ depends on $\epsilon_1$ as shown in \eqref{eq:resampling}.
The SRO is estimated by maximizing this objective function.
As the objective function is usually locally unimodal around the global optimum, the golden section search initialized by a coarse grid search can find the optimal $\epsilon_1$ efficiently.

\section{Proposed Iterative SRO Estimation}

In this section, we propose a joint optimization method of all SROs of an arbitrary number of microphones.
The proposed method alternately updates the SROs and SCMs to increase the log-likelihood for~\eqref{eq:gauss}.
In the update of SROs, we maximize an auxiliary function instead of the intractable log-likelihood.

\subsection{Joint Optimization Problem of all SROs}

Most of the existing SRO estimation methods focused on synchronizing two microphones \cite{cd1,cd2,cd3,cm2,ml1,ml2}.
A na{\"i}ve extension of these methods to the case $M>2$ is the pairwise method as illustrated in Fig.~\ref{fig:overview}-(a).
This method considers the signals observed by the $0$th and $m \neq 0$th microphones as two-channel signals.
Then, the SRO of the $m$th microphone is estimated independently, where the relationship among signals observed by the non-reference microphones is not considered.
To improve the performance of SRO estimation, the relationships among all observed signals should be exploited regardless of the reference microphone as depicted in Fig.~\ref{fig:overview}-(b).

To this end, we propose to jointly optimize all SROs in a maximum likelihood manner.
By denoting the SROs as $\bm{\epsilon} = [\epsilon_0, \ldots, \epsilon_{M-1}]^\mathsf{T}$, the log-likelihood of the SROs and SCMs for the multivariate complex Gaussian model in \eqref{eq:gauss} is given by
\begin{align}
    &\mathcal{L}(\bm{\epsilon}, \bm{V}[0], \ldots, \bm{V}[F-1]) \nonumber \\
    & = \sum_{f=0}^{F-1} \sum_{t=0}^{T-1}
    - \log \det (\bm{V}[f])
    - \widehat{\bm{x}}^{\mathsf{H}}[t,f] \bm{V}^{-1}[f] \widehat{\bm{x}}[t,f],
    \label{eq:likelihood}
\end{align}
where $(\cdot)^\mathsf{H}$ is the Hermitian transpose, and a constant term is omitted.
This log-likelihood is difficult to maximize with respect to the SROs in a closed form.
Furthermore, the golden section search used in the existing method is not applicable to the maximization of such a multivariate function.

\subsection{Alternative Updates of SROs and SCMs}

To maximize \eqref{eq:likelihood} with respect to $\bm{\epsilon}$ and $\bm{V}[f]$, we develop an iterative algorithm that alternately updates them.
For fixed SROs, the SCMs that maximize the log-likelihood in \eqref{eq:likelihood} are easily obtained as
\begin{equation}
    \bm{V}[f] \leftarrow \frac{1}{T} \sum_{t=0}^{T-1} \widehat{\bm{x}}[t,f]
    \widehat{\bm{x}}^{\mathsf{H}}[t,f]. \label{eq:ml-scm}
\end{equation}
Meanwhile, it is difficult to maximize the log-likelihood with respect to $\bm{\epsilon}$ even with fixed $\bm{V}[f]$.
Hence, we use the auxiliary function method~\cite{aux1,aux2} that can handle multivariate non-convex optimization problems and has achieved promising results in array signal processing~\cite{auxiva,auxtde,auxdxcp}.
To be specific, we consider the following objective function by removing a term independent of $\bm{\epsilon}$ from \eqref{eq:likelihood}:
\begin{equation}
    \mathcal{J}(\bm{\epsilon}) = - \sum_{f=0}^{F-1} \sum_{t=0}^{T-1} \widehat{\bm{x}}^{\mathsf{H}}[t,f] \bm{V}^{-1}[f] \widehat{\bm{x}}[t,f].
    \label{eq:objective}
\end{equation}
Then, we introduce an auxiliary variable $\widetilde{\bm{\epsilon}}$ and derive an auxiliary function ${\mathcal{Q}}(\bm{\epsilon} \mid \widetilde{\bm{\epsilon}})$ that satisfies the following properties:
\begin{itemize}
    \item For all $\bm{\epsilon}$ and $\widetilde{\bm{\epsilon}}$, $ \mathcal{J}(\bm{\epsilon}) \geq {\mathcal{Q}}(\bm{\epsilon} \mid \widetilde{\bm{\epsilon}})$.
    \item For all $\bm{\epsilon}$, ${\mathcal{Q}}(\bm{\epsilon} \mid \bm{\epsilon}) = \mathcal{J}(\bm{\epsilon})$.
\end{itemize}
The auxiliary function method alternately updates $\bm{\epsilon}$ and $\widetilde{\bm{\epsilon}}$ to maximize ${\mathcal{Q}}(\bm{\epsilon} \mid \widetilde{\bm{\epsilon}})$.
Thanks to the properties of the auxiliary function, $\argmax_{\widetilde{\bm{\epsilon}}} {\mathcal{Q}}(\bm{\epsilon} \mid \widetilde{\bm{\epsilon}})$ is obtained as $\widetilde{\bm{\epsilon}} \leftarrow \bm{\epsilon}$.
Meanwhile, it depends on the auxiliary function whether $\argmax_{\bm{\epsilon}} {\mathcal{Q}}(\bm{\epsilon} \mid \widetilde{\bm{\epsilon}})$ is obtained in a closed form or not.
The detail of the proposed auxiliary function is explained in the next subsection.

\subsection{Auxiliary Function for Jointly Updating SROs}

To derive the auxiliary function ${\mathcal{Q}}(\bm{\epsilon} \mid \widetilde{\bm{\epsilon}})$, we use the following tractable lower bound of the negative cosine function.
Let $\alpha \in \mathbb{R}_+$, $\beta \in \mathbb{R}$, $\gamma \in \mathbb{R}$, $\theta \in \mathbb{R}$, and $\widetilde{\theta} \in \mathbb{R}$. 
Then, the following inequality holds%
\footnote{
Although the original paper~\cite{auxtde} derived an upper bound of a negative cosine function, the lower bound in \eqref{eq:cos-aux} can be derived in a similar manner.
We thus omit the detailed proof of the inequality.
}%
\cite{auxtde}:
\begin{equation}
  - \alpha \cos(\beta \theta + \gamma) \geq 
  -\frac{\alpha}{2} \sinc(\beta \widetilde{\theta} - \phi) (\beta \theta - \phi)^2 + \eta,
  \label{eq:cos-aux}
\end{equation}
where $\sinc(\cdot) = \sin(x)/x$ if $x\neq0$ and $1$ otherwise, and
\begin{align}
  \phi &= 2 \pi \left\lfloor \frac{\beta \widetilde{\theta} + \gamma}{2\pi} \right\rfloor + \pi -\gamma, \label{eq:mean} \\
  \eta &= \frac{\alpha}{2} \sinc(\beta \widetilde{\theta} -\phi) (\beta \widetilde{\theta} - \phi)^2 - \alpha \cos(\beta \widetilde{\theta} + \gamma). \label{eq:bias}
\end{align}
The equality in \eqref{eq:cos-aux} holds when $\theta = \widetilde{\theta}$.

To reformulate the objective function \eqref{eq:objective} by a sum of the negative cosine functions, we define
\begin{align}
    \bm{\Upsilon}[t,f] = \mathrm{diag}(\bm{x}[t,f])^\mathsf{H} \bm{V}^{-1}[f] \mathrm{diag}({\bm{x}}[t,f]),
\end{align}
where $\mathrm{diag}(\cdot)$ returns a diagonal matrix whose diagonal entries are its input.
By leveraging the conjugate symmetry of $\bm{\Upsilon}[t,f]$ and the Euler's formula~\cite{auxtde2}, the objective function can be reformulated as:
\begin{align}
    \mathcal{J}(\bm{\epsilon}) &= \sum_{t=0}^{T-1} \sum_{f=0}^{F-1} \sum_{m=0}^{M-1} \sum_{n=0}^{M-1} \underline{\mathcal{J}}_{t,f,m,n}(\bm{\epsilon}),
    \label{eq:entry-wise}
    \\
    \underline{\mathcal{J}}_{t,f,m,n}(\bm{\epsilon})
    &=  - |\Upsilon_{m,n}[t,f]| \cos \Bigl(\omega[t,f] (\epsilon_{n} - \epsilon_m) \nonumber \\
    &\hspace{90pt}+ \angle \Upsilon_{m,n}[t,f] \Bigr),
    \label{eq:entry-wise-objective}
\end{align}
where $\omega[t,f] = 2 \pi a t f /F$, and $\angle \cdot $ denotes the principal value of the complex-argument.

Since the entry-wise objective function in \eqref{eq:entry-wise-objective} is a negative cosine function with respect to $\epsilon_{n} - \epsilon_m$, we obtain the following auxiliary function based on \eqref{eq:cos-aux}:
\begin{align}
    \underline{\mathcal{Q}}_{t,f,m,n}(\bm{\epsilon} \mid \widetilde{\bm{\epsilon}}) &= - \lambda_{m,n}[t,f] \Bigl(\omega[t,f] (\epsilon_{n} - \epsilon_m)
    \nonumber \\
    & \hspace{35pt}
    -\mu_{m,n}[t,f]\Bigr)^2
    \! + \nu_{m,n}[t,f], \label{eq:aux-entry}
\end{align}
where $\nu_{m,n}[t,f]$ does not depend on $\bm{\epsilon}$, and 
\begin{align}
    \!\! \xi_{m,n}[t,f] \!&= \omega[t,f] (\widetilde{\epsilon}_{n} - \widetilde{\epsilon}_m),
    \label{eq:xi} \\
    \!\! \lambda_{m,n}[t,f] \!&= \frac{|\Upsilon_{m,n}[t,f]|}{2} \sinc \Bigl(\xi_{m,n}[t,f] - \mu_{m,n}[t,f] \Bigr), \!\!\! \\
    \!\! \mu_{m,n}[t,f] \!&= 2 \pi \left\lfloor \frac{\xi_{m,n}[t,f] + \angle \Upsilon_{m,n}[t,f]}{2\pi} \right\rfloor \nonumber \\
    &  \hspace{90pt} + \pi - \angle \Upsilon_{m,n}[t,f],
    \label{eq:mu}
\end{align}
where $\xi_{m,n}[t,f] - \mu_{m,n}[t,f]$ is in $[-\pi, \pi)$, and thus $\lambda_{m,n}[t,f] \geq 0$.
We stress that \eqref{eq:entry-wise-objective} and \eqref{eq:aux-entry} correspond to the left and right side of \eqref{eq:cos-aux}, respectively.
Finally, we obtain the auxiliary function ${\mathcal{Q}}(\bm{\epsilon} \mid \widetilde{\bm{\epsilon}})$ by summing up $\underline{\mathcal{Q}}_{t,f,m,n}(\bm{\epsilon} \mid \widetilde{\bm{\epsilon}})$ for all T-F bins and microphone pairs.

\begin{algorithm}[t!]
\caption{Iterative Algorithm to Estimate SROs}
\algsetup{indent=2mm}
\begin{algorithmic}
\renewcommand{\algorithmicrequire}{\textbf{Input:}}
\renewcommand{\algorithmicensure}{\textbf{Output:}}
\REQUIRE Initial estimate of SROs $\bm{\epsilon}$, $\bm{D}$, $\bm{u}$, $\omega[t,f]$
\ENSURE Final estimate of SROs $\bm{\epsilon}$
\FOR {$k=0, \ldots, K-1$}
\STATE $\widehat{x}_m[t,f] = x_m[t,f] \exp\left(\frac{2 \pi j a t f \epsilon_m}{F} \right)$
\STATE $\bm{V}[f] \leftarrow (1/T) \sum_{t=0}^{T-1} \widehat{\bm{x}}[t,f] \widehat{\bm{x}}^{\mathsf{H}}[t,f]$
\STATE $\bm{\Upsilon}[t,f] = \mathrm{diag}(\bm{x}[t,f])^\mathsf{H} \bm{V}^{-1}[f] \mathrm{diag}({\bm{x}}[t,f])$
\FOR {$k'= 0, \ldots, K'-1$}
\STATE $\widetilde{\bm{\epsilon}} \leftarrow \bm{\epsilon}$
\STATE $\xi_{m,n}[t,f] \!= \omega[t,f] (\widetilde{\epsilon}_{n} - \widetilde{\epsilon}_m)$
\STATE $\lambda_{m,n}[t,f] \!= \frac{|{\Upsilon}_{m,n}[t,f]|}{2} \sinc \Bigl(\xi_{m,n}[t,f] - \mu_{m,n}[t,f] \Bigr)$
\STATE $\mu_{m,n}[t,f] \!= 2 \pi \!\left\lfloor\! \frac{\xi_{m,n}[t,f] + \angle {\Upsilon}_{m,n}[t,f]}{2\pi} \!\right\rfloor\! + \pi \!-\! \angle {\Upsilon}_{m,n}[t,f]$ \!\!
\STATE $\bm{A} = \sum_{f=0}^{F-1} \sum_{t=0}^{T-1} \omega^2[t,f] \bm{\Lambda}[t,f]$
\STATE $\bm{b} = \sum_{f=0}^{F-1} \sum_{t=0}^{T-1} \omega[t,f] \bm{\Lambda}[t,f] \bm{\mu}[t,f]$
\STATE $\left(
\begin{array}{c}
\bm{\epsilon} \\
\rho
\end{array}
\right) \leftarrow
\left(
\begin{array}{cc}
\bm{D}^\mathsf{T} \bm{A} \bm{D}
& \bm{u} \\
\bm{u}^{\mathsf{T}} & 0
\end{array}
\right)^{-1}
\left(
\begin{array}{c}
\bm{D}^\mathsf{T} \bm{b} \\
0
\end{array}
\right)$
\ENDFOR
\ENDFOR
\end{algorithmic}
\label{alg:prop}
\end{algorithm}

Since the auxiliary function $\mathcal{Q}(\bm{\epsilon} \mid \widetilde{\bm{\epsilon}})$ is a sum of negative quadratic functions, we can easily maximize it under the constraint $\epsilon_0 = 0$.
By considering the Karush--Kuhn--Tucker (KKT) condition, the optimal $\bm{\epsilon}$ is obtained by solving the following linear equation~\cite{boyd}:
\begin{equation}
\left(
\begin{array}{cc}
\bm{D}^\mathsf{T} \bm{A} \bm{D}
& \bm{u} \\
\bm{u}^{\mathsf{T}} & 0
\end{array}
\right)
\left(
\begin{array}{c}
\bm{\epsilon}^\star \\
\rho^\star
\end{array}
\right)
=
\left(
\begin{array}{c}
\bm{D}^\mathsf{T} \bm{b} \\
0
\end{array}
\right),
\label{eq:kkt}
\end{equation}
where $\bm{u} = [1,0,\ldots,0]^\mathsf{T} \in \mathbb{R}^{M}$, $\rho^{\star} \in \mathbb{R}$ is the KKT multiplier, $\bm{D} \in \mathbb{R}^{M^2 \times M}$ is a matrix that computes the difference in $\bm{\epsilon}$ as
\begin{equation}
     \epsilon_n - \epsilon_m = (\bm{D}\bm{\epsilon})_{mM+n},
\end{equation}
and $\bm{A} \in \mathbb{R}^{M^2 \times M^2}$ and $\bm{b} \in \mathbb{R}^{M^2}$ are given by
\begin{align}
    \bm{A} &= \sum_{f=0}^{F-1} \sum_{t=0}^{T-1} \omega^2[t,f] \bm{\Lambda}[t,f], \label{eq:kkt-mat} \\
    \bm{b} &= \sum_{f=0}^{F-1} \sum_{t=0}^{T-1} \omega[t,f] \bm{\Lambda}[t,f] \bm{\mu}[t,f]. \label{eq:kkt-vec}
\end{align}
Here, $\bm{\Lambda}[t,f]$ is a diagonal matrix whose $(mM+n, mM+n)$th entry is given by $\lambda_{m,n}[t,f]$, and the $(mM+n)$th entry of $\bm{\mu}[t,f]$ is given by $\mu_{m,n}[t,f]$.
In each inner iteration, $\bm{\epsilon}$ is updated to maximize the auxiliary function ${\mathcal{Q}}(\bm{\epsilon} \mid \widetilde{\bm{\epsilon}})$ as follows:
\begin{equation}
\left(
\begin{array}{c}
\bm{\epsilon} \\
\rho
\end{array}
\right) \leftarrow
\left(
\begin{array}{cc}
\bm{D}^\mathsf{T} \bm{A} \bm{D}
& \bm{u} \\
\bm{u}^{\mathsf{T}} & 0
\end{array}
\right)^{-1}
\left(
\begin{array}{c}
\bm{D}^\mathsf{T} \bm{b} \\
0
\end{array}
\right).
\label{eq:kkt-answer}
\end{equation}

The proposed algorithm is summarized in Algorithm~\ref{alg:prop}, where $k=0, \ldots,K-1$ and $k' = 0, \ldots, K'-1$ are the iteration counters.
In each outer iteration, the auxiliary function method is used to update the SROs $K'$ times.
Owing to the property of the auxiliary function ${\mathcal{Q}}(\bm{\epsilon} \mid \widetilde{\bm{\epsilon}})$, this algorithm ensures that the log-likelihood $\mathcal{L}(\bm{\epsilon}, \bm{V}[0], \ldots, \bm{V}[F-1])$ does not decrease.
When $M=2$, the proposed method aims to maximize the same objective function considered in the existing methods~\cite{ml1,ml2}.
When $M>2$, the proposed method can consider the consistency of the estimated SROs based on the entire relationship among signals observed by the non-reference microphones.
On the other hand, the pairwise method cannot leverage the relationships due to their separate optimization.

\section{Experimental Evaluations}

In this section, we evaluated the proposed method on multichannel speech mixtures that imitate meeting recordings with some speakers.
We first investigated the convergence of the proposed method.
To visualize the log-likelihood, we used a two-channel speech signal where the log-likelihood can be reformulated as the univariate function with respect to $\epsilon_1$.
Second, the effectiveness of the joint optimization of all SROs was validated on the four-channel speech mixtures.

\begin{figure}[t]
\centering
\includegraphics[width=0.96\columnwidth]{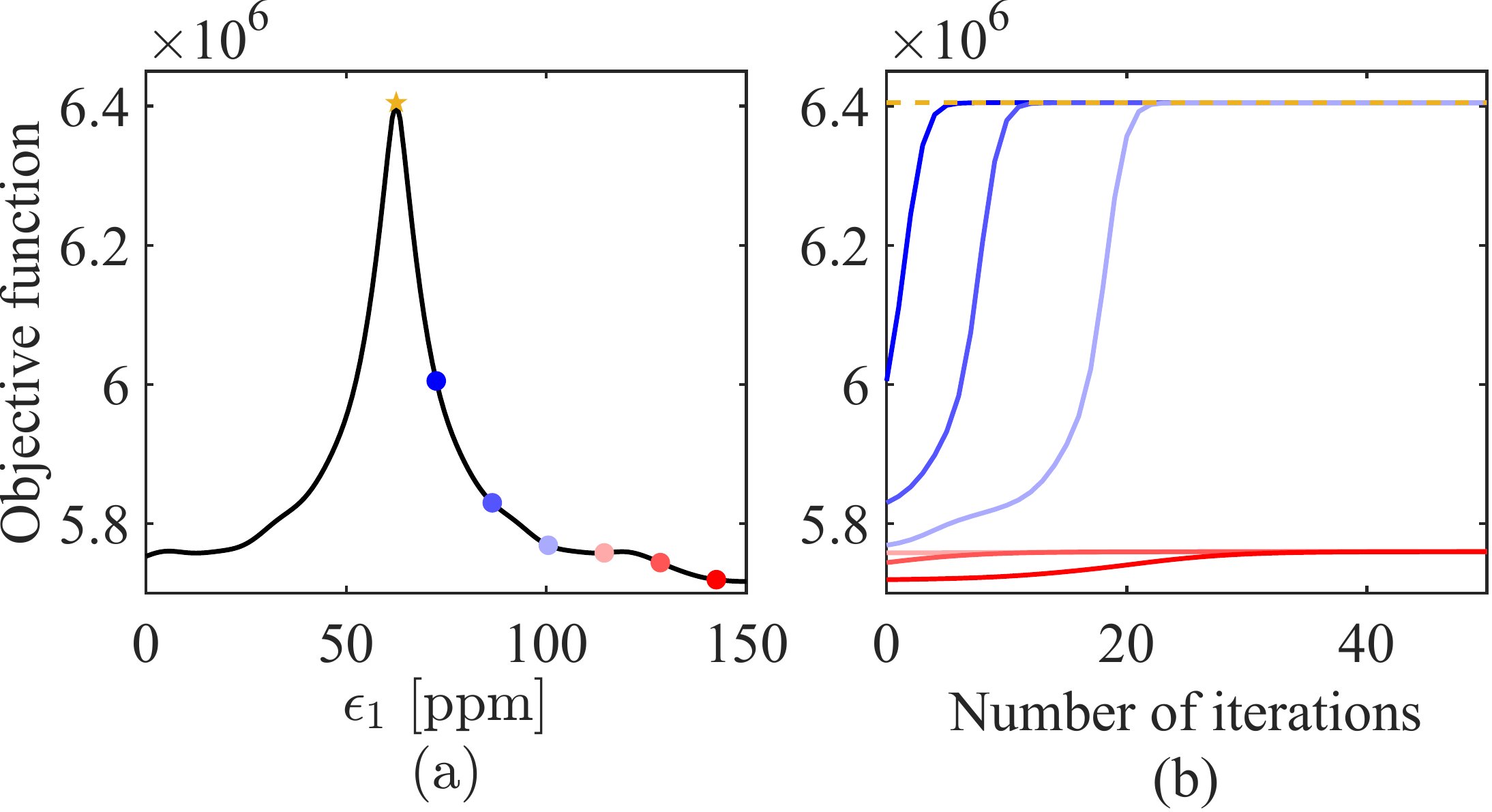}
\caption{(a) Objective function with respect to $\epsilon_1$ in parts per million (ppm).
(b) Objective function with respect to number of iterations.
Each solid line uses $\epsilon_1$ at the circle of the same color in (a) as the initial value.
The yellow star and dotted lines correspond to the oracle $\epsilon_1$.
}
\label{fig:convergence}
\end{figure}

\subsection{Convergence of Proposed Method}

In this experiment, we used a two-channel signal with a single speaker.
The source signal of $10$ s length was generated by concatenating utterances in the Voice Conversion Challenge (VCC) $2018$ dataset~\cite{vcc2018}.
The source signal was downsampled to $16000$ Hz.
To synthesize the reverberant signal, we performed room simulations using the {\tt pyroomacoustics} toolbox~\cite{pyra}.
The sound source and microphones were randomly located in a room of $6.0$ m $\times$ $8.0$ m $\times$ $4.0$ m size.
The reverberation time was also randomly sampled from $[0.2, 0.4]$ s.
The signal measured by the non-reference microphone was further resampled at $16001$ Hz.
For the STFT, the $2048$-point-long Hann window was used with $1024$-point shifts, where the number of DFT points was $4096$.
In Algorithm~\ref{alg:prop}, $K'$ was $1$.

Fig.~\ref{fig:convergence}-(a) shows the log-likelihood in~\eqref{eq:likelihood} as a function of $\epsilon_1$, where the SCMs were calculated by~\eqref{eq:ml-scm}.
The yellow star corresponds to the oracle $\epsilon_1$, and multiple initial values of the proposed method are depicted by circles.
Fig.~\ref{fig:convergence}-(b) shows the convergence of the proposed method with different initial values.
According to Fig~\ref{fig:convergence}-(a), the log-likelihood can be viewed as a unimodal function around the oracle $\epsilon_1$.
As a result, the proposed method converged close to the oracle SRO when the initial value was appropriate.
Even when the initial value was outside of the appropriate interval, the proposed method converged to a local maximum as theoretically guaranteed by the property of the auxiliary function method. 

\begin{figure}[t]
\centering
\includegraphics[width=0.99\columnwidth]{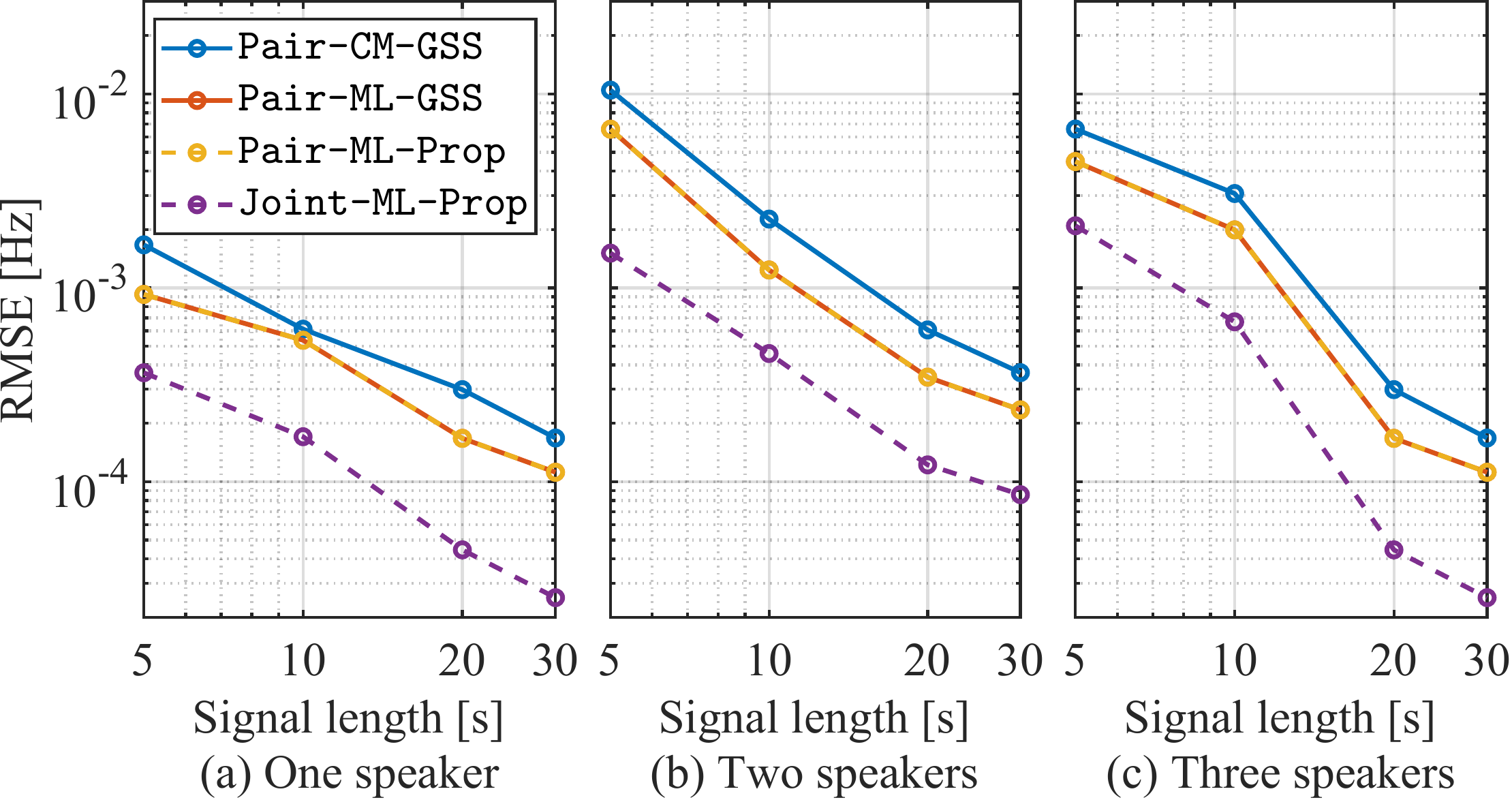}
\caption{
RMSEs of the estimated sampling rates for (a) one-speaker signals, (b) two-speaker mixtures, (c) three-speaker mixtures observed by four microphones.
}
\label{fig:sro-mch}
\end{figure}

\subsection{Synchronization of More Than Two Microphones}

To confirm the effectiveness of the joint optimization of all SROs, we evaluated the performance of SRO estimation using four distributed microphones.
In addition to the one-speaker signals, we synthesized two and three-speaker mixtures. 
In each case, $10$ signals of $30$ s length were synthesized.
The signals measured by non-reference microphones were randomly resampled at $r_m \in [15999, 16001]$ Hz.
Other conditions were the same as in the previous experiment.

The proposed method (\texttt{Joint-ML-Prop}) was compared with the pairwise maximum likelihood estimation method by the golden section search (\texttt{Pair-ML-GSS})~\cite{ml1} and by Algorithm~\ref{alg:prop} (\texttt{Pair-ML-Prop}).
We also investigated the performance of the pairwise correlation maximization method (\texttt{Pair-CM-GSS})~\cite{cm2}.
For all methods, we initialized SROs by a coarse grid search in a pairwise manner.
The search range was from $-100$ ppm to $100$ ppm with $100$ grids, which is finer than the grids in Fig.~\ref{fig:convergence}.
We expect that this initialization enabled us to avoid bad local optima.
Then, the initial estimate was refined by the golden section search or Algorithm~\ref{alg:prop}.

Fig.~\ref{fig:sro-mch} shows the root mean square errors (RMSEs) of the estimated sampling rates for different signal lengths.
\texttt{Pair-ML-GSS} and \texttt{Pair-ML-Prop}  resulted in the same RMSE.
That is, the difference of the optimization algorithms did not affect the performance in our experimental conditions. 
Meanwhile, \texttt{Joint-ML-Prop} outperformed all pairwise methods regardless of the number of speakers.
This result confirmed the effectiveness of leveraging the relationships between all microphone pairs and optimizing all the SROs jointly.
In all conditions, \texttt{Joint-ML-Prop} with $5$-second-long signals was comparable to \texttt{Pair-ML-Prop} with $10$-second-long signals.
This result indicates that the proposed method can perform well with shorter signals, which is desirable to adopt to time-varying SROs~\cite{tvsro} and unstationary environments~\cite{tvscm}.

\section{Conclusion}

In this paper, we propose a joint optimization method for all the SROs of multiple devices.
The proposed method is based on the probabilistic model of the entire multichannel signal and estimates the SROs in a maximum likelihood manner.
As the key idea, we maximize a tractable auxiliary function with respect to the SROs instead of the log-likelihood itself.
Experimental results confirmed the effectiveness of the proposed joint optimization method compared with the pairwise methods.
Future work includes an investigation of the robustness of the proposed method in real environments.

\section{Acknowledgment}
This work was supported by JSPS KAKENHI Grant Numbers JP20H00613 and JP21J21371, and JST CREST Grant Number JPMJCR19A3, Japan.

\clearpage

\end{document}